\documentclass[10pt,draft]{dis03}
\usepackage{epsf,amsmath}

\begin{document}

\title{RECENT RESULTS IN THE STATISTICAL APPROACH \\
 OF POLARIZED PDF AND EXTENSION TO \\
 BARYON FRAGMENTATION FUNCTIONS 
\thanks{Talk presented at DIS2003, XI International Workshop on Deep Inelastic Scattering,
St. Petersburg, 23-27 April,2003}}

\author{JACQUES SOFFER \\
Centre de Physique Th\'eorique, CNRS Luminy Case 907, \\
13288 Marseille Cedex 09, France \\
E-mail: soffer@cpt.univ-mrs.fr }

\maketitle

\begin{abstract}
\noindent  Parton distributions are
constructed in a statistical physical picture of the nucleon. 
The chiral properties of QCD lead to strong relations between quarks and 
antiquarks distributions and the importance of the Pauli exclusion principle
is also emphasized. A global next-to-leading order QCD analysis of unpolarized and polarized
deep-inelastic scattering data is performed to determine a small number of
free parameters. Some predictions are compared to recent experimental results. 
We also present an extension of the statistical approach to the description of baryon
fragmentation functions.

\end{abstract}

Deep-inelastic scattering (DIS) of leptons on hadrons has been extensively
studied, over the last twenty years or so, both theoretically and experimentally, to
extract the polarized parton distributions functions (PDF) of the nucleon.
As it is well known, the unpolarized light quarks ($u,d$) distributions are fairly 
well determined. Moreover, the data exhibit a clear evidence for a
flavor-asymmetric light sea, {\it i.e.} $\bar d > \bar u$, which can be
understood in terms of the Pauli exclusion principle, based on the fact 
that the proton contains two $u$ quarks and only one $d$ quark. 
Larger uncertainties still persist for the gluon ($G$) and the heavy quarks 
($s,c$) distributions. The corresponding polarized gluon and $s$ quark
distributions ($\Delta G, \Delta s$) are badly constrained and we just begin to
uncover a flavor asymmetry, for the corresponding polarized light sea, namely  
$\Delta \bar u \neq \Delta \bar d$. 
Whereas the signs of the polarized light quarks distributions are 
essentially well established, $\Delta u > 0$ and $\Delta d < 0$, 
this is not the case for $\Delta \bar u $ and $\Delta \bar d $.
In this report, essentially based on Refs.\cite{BBS,BBS1}, we briefly recall how 
we construct a complete set of polarized parton
(all flavor quarks, antiquarks and gluon) distributions. 
Our motivation is to use the statistical approach to
build up : $q_i$, $\Delta q_i$, $\bar q_i$, $\Delta \bar q_i$, $G$ and $\Delta
G$, in terms of a very small number of free parameters. 
A flavor separation for the unpolarized and polarized light sea
is automatically achieved in a way dictated by our approach.

The existence of the correlation, broader shape higher first moment, suggested
by the Pauli principle, has inspired the introduction of Fermi-Dirac (Bose-Einstein)
functions for the quark (gluon) distributions \cite{bour96}. After many years
of research, we recently proposed \cite{BBS}, at the input scale
$Q_0^2 = 4 \mbox{GeV}^2$

\begin{eqnarray}
x u^{+}(x,Q^2_0) &=& {AX_{0u}^{+} x^b \over \exp[(x-X_{0u}^{+})/{\bar x}]
+1} + {\tilde A x^{\tilde b} \over \exp(x/{\bar x}) +1}\, , 
\label{eq5} \\
x \bar u^{-}(x,Q^2_0) &=& {\bar A (X_{0u}^{+})^{-1}x^{2b} \over 
\exp[(x+X_{0u}^{+})/{\bar x}]+1} + {\tilde A x^{\tilde b} \over 
\exp(x / {\bar x}) +1} \, , \label{eq6}\\
x G(x,Q^2_0) &=& {A_G x^{\tilde b +1} \over \exp(x /{\bar x}) - 1} \label{eq7}
\, ,
\end{eqnarray}
and similar expressions for the other light quarks ($u^{-},
d^{+}~\mbox{and}~ d^{-}$) and their antiparticles.
We assumed $\Delta G(x,Q^2_0) = 0$ and the strange parton distributions 
$s(x,Q^2_0)$ and $\Delta s(x,Q^2_0)$ are simply related \cite{BBS} to
$\bar q(x,Q^2_0)$ and $\Delta \bar q(x,Q^2_0)$, for $q = u,d$.
A peculiar aspect of this approach, is that it solves the problem of
desentangling the $q$ and $\bar q$ contribution through the relationship 
\cite{BBS}
\begin{equation}
  X_{0u}^{+} + X_{0 \bar u}^{-} = 0 \, ,
\label{eq11}
\end{equation}
and the corresponding one for the other light quarks and their antiparticles.
It allows to get the $\bar q(x)$ and $\Delta \bar q(x)$ distributions
from the ones for $q(x)$ and $\Delta q(x)$.
 
 By performing a next-to-leading order QCD evolution of these parton distributions, we
were able to obtain a good description of
a large set of very precise data on $F_2^p(x,Q^2), F_2^n(x,Q^2), xF_3^{\nu N}(x,Q^2)$
and $g_1^{p, d, n}(x,Q^2)$ data, in correspondance with the {\it eight} free parameters.
Therefore crucial tests will be provided by measuring flavor and spin 
asymmetries for antiquarks, for which we expect \cite{BBS} 
\begin{equation}
\Delta \bar u(x) > 0 > \Delta \bar d(x) \, ,
\label{eq12}
\end{equation}
\begin{equation}
\Delta \bar u(x) - \Delta \bar d(x) \simeq \bar d(x) - \bar u(x) > 0 \, .
\label{eq13}
\end{equation}
The inequality $\bar d(x) - \bar u(x) > 0$ has the right sign to agree with the
defect in the Gottfried sum rule \cite{got67} and if Eq. (6) above is satisfied,
it means that the antiquark polarization contributes to about $15\%$ to the
Bjorken sum rule \cite{bj}, which is not negligible. For illustration, we show in Figs. 1 and 2, the predictions
of the statistical approach with recent DIS results.

%\clearpage
%\newpage
\begin{figure}%[htb]
\vspace*{-2.0cm}
\begin{center}
\leavevmode {\epsfxsize= 6.5cm \epsffile{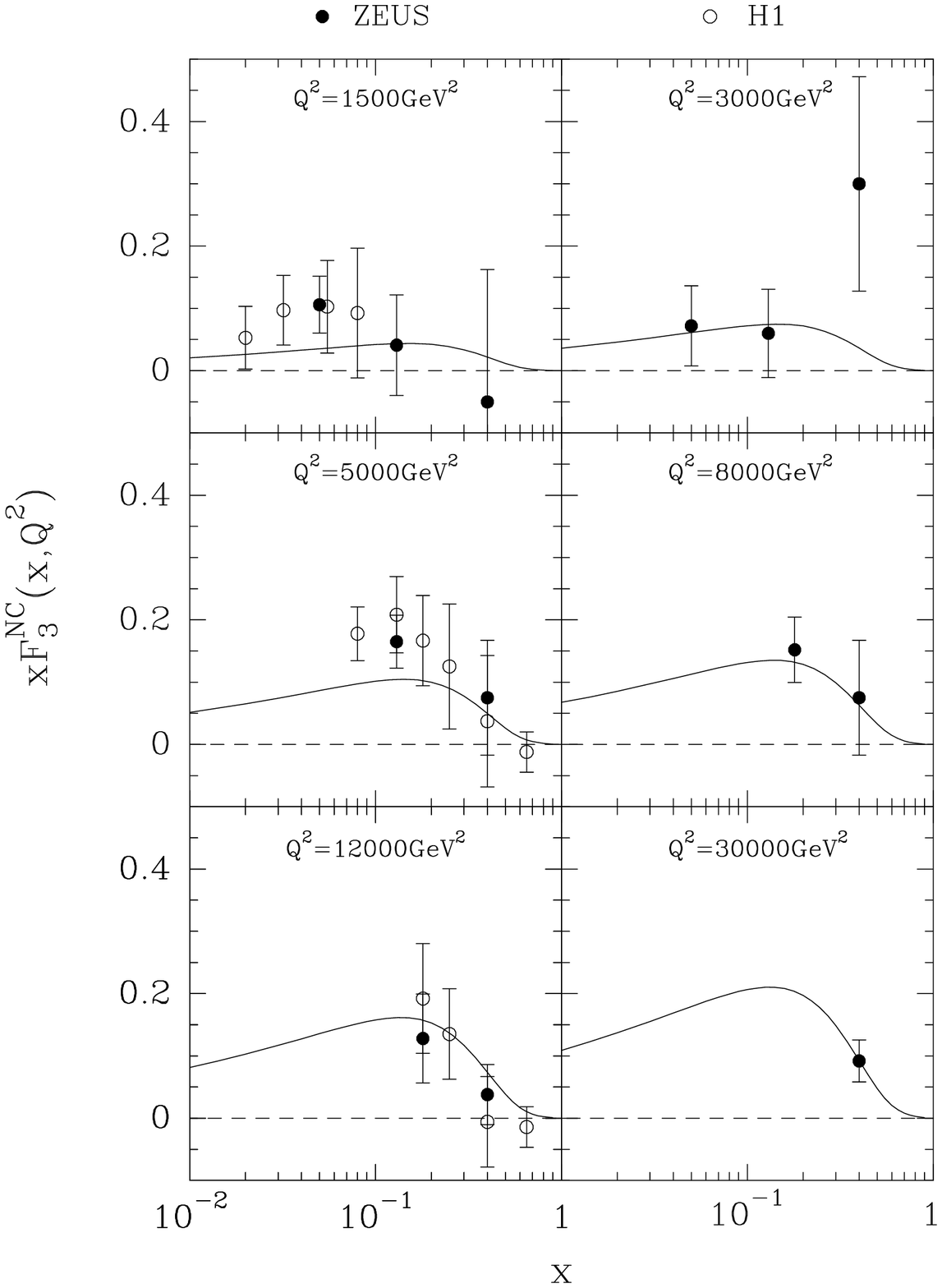}}
%\end{center}
%\special{psfile=xF3.eps voffset=-10 vscale=40
%hscale= 40 hoffset=20 angle=0}
%\centerline{\epsfxsize=3.3in\epsfbox{xF3.eps}}
\caption[*]{\baselineskip 1pt
 The structure function $xF_3^{NC}$ as a function of $x$, for different $Q^2$.
%\caption{ The structure function $xF_3^{NC}$ as a function of $x$, for different $Q^2$.
Data from ZEUS Coll. \cite{zeus02}, H1 Coll. \cite{h101} and the curves are the predictions
from the statistical approach (Taken from Ref. \cite{BBS1}).}
%\end{center}
%\end{figure}
%\vspace*{4.5cm}
%\begin{figure}%[htb]
%\vspace*{4.5cm}
%\begin{center}
\leavevmode {\epsfxsize= 6.cm \epsffile{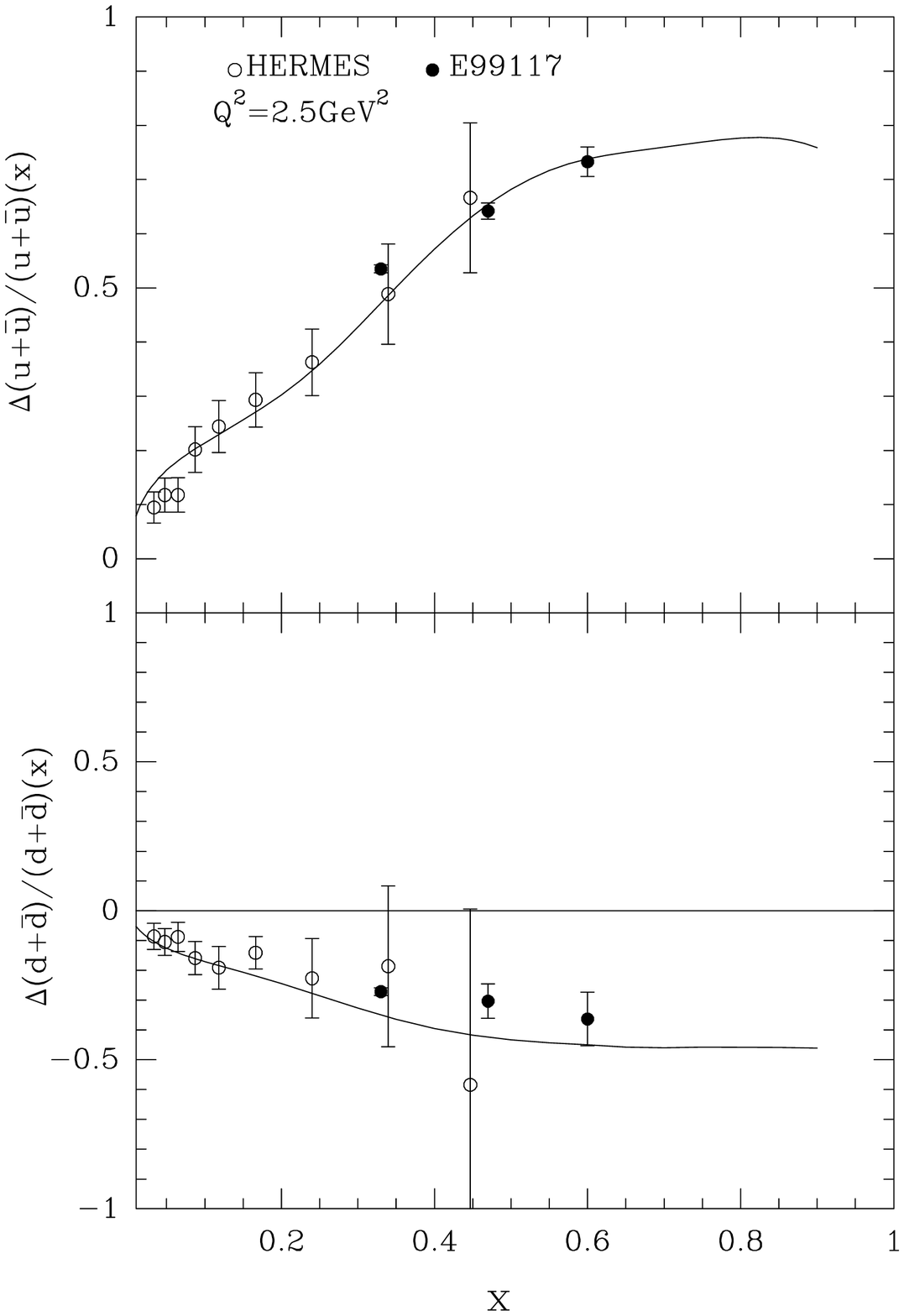}}
%\end{center}
%\special{psfile=jeflab.eps voffset=-10 vscale=30
%hscale= 30 hoffset=25 angle=0}
%\centerline{\epsfxsize=2.9in\epsfysize=3.1in\epsfbox{jeflab.eps}}
\caption[*]{\baselineskip 1pt
 Results for $\Delta(u + \bar u)/(u + \bar u)\,(x)$
% \caption{ Results for $\Delta(u + \bar u)/(u + \bar u)\,(x)$
and $\Delta(d + \bar d)/(d + \bar d)\,(x)$ from Ref. \cite{jlab02},  
compared to the statistical model predictions.}
\end{center}
\end{figure}

\clearpage

\begin{figure}[!thb]
\vspace*{-2.0cm}
\begin{center}
%\special{psfile=fig4.eps voffset=-50 vscale=50
%hscale= 50 hoffset=20 angle=0}
\centerline{\epsfxsize=3.5in\epsfbox{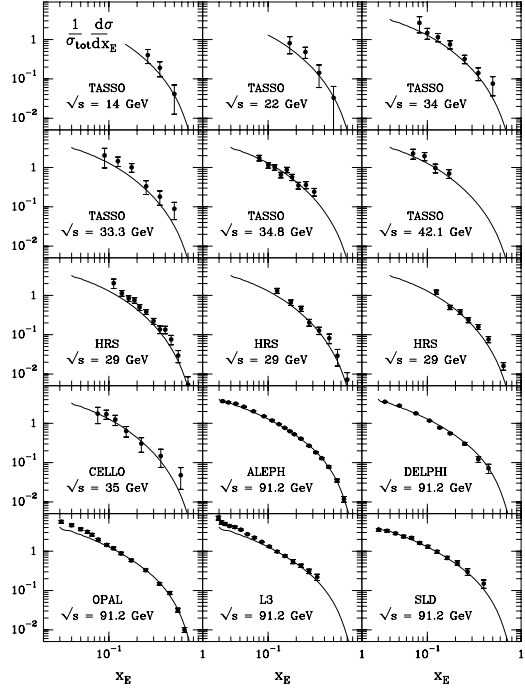}}
\caption[*]{ for $\Lambda$ production in $e^+e^-$ annihilation
at several energies, as a function of $x_E$ (Taken from Ref. \cite{BS}).}
\end{center}
\end{figure}

We now briefly discuss an extension of this approach to construct the octet baryons 
fragmentation functions (FF).
The characteristics of the model are determined by using some data
on the inclusive production of proton and $\Lambda$ in unpolarized deep 
inelastic scattering and a next-to-leading analysis of the available 
experimental data on the production of unpolarized octet baryons 
in $e^+e^-$ annihilation.
We believe that the statistical features which were proposed
to build up the nucleon PDF in Ref.~\cite{BBS}, can be used also 
to construct the FF for the octet baryons.
We assume that the parton ($p$) to hadron ($h$) FF $D_{p}^h(x)$, 
at an input energy scale $Q_0^2$, is proportional to

\begin{equation}
[\exp[(x - X_{0})/{\bar x}] \pm 1]^{-1}~,
\label{1}
\end{equation}
where the {\it plus} sign for quarks and antiquarks, corresponds to a 
Fermi-Dirac distribution and the {\it minus} sign for gluons, corresponds 
to a Bose-Einstein distribution. Here $X_{0}$ is a constant which plays 
the role of the {\it thermodynamical potential} of the quark 
hadronization into a hadron and $\bar x$ is the 
{\it universal temperature}, which is assumed to be equal 
for all octet baryons and it is reasonable to take its value to be the same
as for the nucleon PDF. For more details the interested reader can go to
Ref. \cite{BS} and here we just show our results for  $\Lambda$ production, displayed in Fig. 3,
for the energies $\sqrt{s}=14, 22, 29, 33.3, 34.8, 42.1, 91.2~\rm{GeV}$.
It is clear from Fig. 3, which covers a
sizeable energy domain, that the scaling violations in this range are very 
small.
Our results show that both parton distributions and fragmentation 
functions are compatible with the statistical approach, in terms of a few free
parameters.

%\clearpage

\section*{Acknowledgements} The author is thankful to INTAS (Project
587, call 2000), which provided the financial support for his attendance
to DIS2003.

\end{document}